\documentclass{article}

\PassOptionsToPackage{numbers, compress, square}{natbib}

\usepackage[final]{neurips_2024_ml4ps}
\usepackage[pdftex]{graphicx}
\usepackage{subcaption}
\usepackage{amsmath,amssymb}
\usepackage{xspace}
\usepackage{multirow}
\usepackage{amsmath}  
\usepackage{bm}  
\usepackage{pifont}
\usepackage{booktabs} 

\newcommand{\pt}{\ensuremath{p_{\text{T}}}\xspace}
\newcommand{\jetclass}{{\textsc{JetClass}}\xspace}
\newcommand{\rej}[1]{\ensuremath{\text{Rej}_{#1}}\xspace}

\newcommand{\hbb}{\ensuremath{H\to b \bar{b}}\xspace}
\newcommand{\hcc}{\ensuremath{H\to c \bar{c}}\xspace}
\newcommand{\hgg}{\ensuremath{H\to g g}\xspace}

\newcommand{\hqqqq}{\ensuremath{H\to 4 q}\xspace}
\newcommand{\hlvqq}{\ensuremath{H\to \ell \nu q q'}\xspace}
\newcommand{\tbqq}{\ensuremath{t\to b q q'}\xspace}
\newcommand{\tblv}{\ensuremath{t\to b \ell \nu}\xspace}
\newcommand{\wqq}{\ensuremath{W\to q q'}\xspace}
\newcommand{\zqq}{\ensuremath{Z\to q \bar{q}}\xspace}
\newcommand{\qgj}{\ensuremath{q/g}\xspace}

\usepackage[utf8]{inputenc} 
\usepackage[T1]{fontenc}    
\usepackage{hyperref}       
\usepackage{url}            
\usepackage{booktabs}       
\usepackage{amsfonts}       
\usepackage{nicefrac}       
\usepackage{microtype}      
\usepackage{xcolor}         

\title{Interpreting Transformers for Jet Tagging}

\author{%
  Aaron Wang\\
  University of Illinois Chicago\\
  Chicago, IL 60607\\
  \texttt{aaronw5@uic.edu}\\
  \And
  Abhijith Gandrakota \quad Jennifer Ngadiuba\\
  Fermi National Accelerator Laboratory\\
  Batavia, IL 60510\\
  \texttt{\{abhijith,ngadiuba\}@fnal.gov}
  \And
  Vivekanand Sahu \quad Priyansh Bhatnagar \quad Elham E Khoda \quad Javier Duarte\\
  University of California San Diego\\
  La Jolla, CA 92093\\
  \texttt{\{vsahu,prbhatnagar,ekhoda,jduarte\}@ucsd.edu}\\
}

\begin{document}
\begin{flushright}
FERMILAB-CONF-24-0868-CMS-LDRD
\end{flushright}

\maketitle

\begin{abstract}
Machine learning (ML) algorithms, particularly attention-based transformer models, have become indispensable for analyzing the vast data generated by particle physics experiments like ATLAS and CMS at the CERN LHC.
Particle Transformer (ParT), a state-of-the-art model, leverages particle-level attention to improve jet-tagging tasks, which are critical for identifying particles resulting from proton collisions.
This study focuses on interpreting ParT by analyzing attention heat maps and particle-pair correlations on the $\eta$-$\phi$ plane, revealing a binary attention pattern where each particle attends to at most one other particle.
At the same time, we observe that ParT shows varying focus on important particles and subjets depending on decay, indicating that the model learns traditional jet substructure observables.
These insights enhance our understanding of the model's internal workings and learning process, offering potential avenues for improving the efficiency of transformer architectures in future high-energy physics applications.
\end{abstract}

\vspace{-0.5em}
\section{Introduction}
\vspace{-0.5em}

Machine learning (ML) algorithms are becoming crucial for effectively analyzing the enormous data produced by particle physics experiments like ATLAS and CMS at the CERN LHC~\cite{Harris:2022qtm}.
Attention-based transformer models~\cite{NIPS2017_3f5ee243}, with their ability to effectively capture and weigh the relative importance of different elements in input data, have revolutionized various domains, including natural language processing and computer vision~\cite{ramesh2022hierarchicaltextconditionalimagegeneration, geminiteam2024geminifamilyhighlycapable, brown2020languagemodelsfewshotlearners}.
The ability of the attention mechanism to discern intricate correlations in the collimated spray of particles jets initiated from massive particle decays has proven invaluable in searches for new physics and standard model measurements~\cite{CMS-PAS-HIG-23-012}. 

High-energy collisions of protons at LHC can create new unstable particles which then decay and produce sprays of outgoing particles, called jets, which are observed by these experiments\cite{Andersson:1987pr}.
Jet tagging, or the process of identifying the particle that initiates this spray, is a critical step in data analysis at the LHC~\cite{Qu_2020,Liu:2023dio}.
One such state-of-the-art model for jet tagging is Particle Transformer (ParT)~\cite{Qu:2022mxj}, which is a transformer variant that leverages physics-inspired particle-particle pairwise features (or ``interactions'') to augment the attention mechanism.
Although this is one of the best performing jet-tagging models, there is effort required to interpret the underlying workflow and why the model performs so well.

The attention-based architecture of transformers may offer deeper insight into the inner workings of neural networks~\cite{Chefer_2021_CVPR}.
In the case of a jet tagger, the attention matrix captures the particle-to-particle correlations potentially making these ML models more interpretable.
These attention scores between particles highlight the most important connections between particles necessary for classification.
This helps us to understand whether the neural network is learning the physics we know.
In addition, looking at which parts of the input the neural network focuses on can help us improve the computational performance by pruning unnecessary parts of the neural network architecture, resulting in a more efficient jet tagging model.  
\paragraph{Related Work}
This paper builds on prior research that has applied machine learning, specifically graph neural networks (GNNs), to particle physics problems at the CERN LHC~\cite{Shlomi:2020gdn,Duarte:2020ngm}. 
Reference~\cite{mokhtar2022graphneuralnetworkslearn} attempts to explain how ParticleNet~\cite{Qu_2020}, a GNN designed for jet tagging learns to classify three-prong hadronic top quark decays using edge relevancy ($R$) graphs based on layerwise relevance propagation.
Similarly, we visualize the attention scores between pairs of particles in order to investigate how ParT classifies jets.

\vspace{-0.5em}
\section{Methods}
\vspace{-0.5em}
This study uses the pre-trained ParT model, trained using the \jetclass dataset~\cite{Qu:2022mxj,JetClass}, which consists of 100 million examples of jets for training, 5 million for validation, and 20 million for testing, containing 10 classes of jets with different decay modes of quarks, gluons, $W$ and $Z$ bosons, Higgs bosons, and top quarks.
The classes are \qgj, \hbb, \hcc, \hgg,  \hqqqq, \hlvqq, \tbqq, \tblv, \wqq, and \zqq, indicating the decays to combination of light quarks ($q$), charm quark ($c$), bottom quark ($b$), and lepton ($\ell$).
There are 17 input features divided into 3 categories: kinematics, particle ID, and trajectory displacement.
Each jet is composed of a maximum of 128 particles, ordered by \pt, with an average of 30--50 particles per jet.



ParT introduces a modified attention mechanism, called particle multihead attention (P-MHA).
Given a particle-level representation of a jet with $N$ constituents with each constituent having $d$ features, $x \in \mathbb R^{N\times d}$, the attention is computed as
\begin{align}
\begin{split}
 \text{P-MHA}(x) &= \text{concat}(\text{head}_1, \dots, \text{head}_h)W^O \\
 \text{head}_i  &=  \text{softmax}\left(\frac{xW_i^W(xW_i^K)^{\intercal}}{\sqrt{d_k}} + U_i\right)xW_i^V,
 \end{split}
\end{align}
where $U_i\in \mathbb{R}^{N\times N}$ is a single feature for each pair of particles, which can be concatenated into a larger structure, known as the pairwise interaction matrix $U \in \mathbb{R}^{N\times N \times d'}$.

In ParT, the number of pairwise features $d'$ must be equal to the number of heads $h$, which is chosen to be 8.
In addition, $d_k = 16$ so that the overall output dimension is the same as the input dimension $d = hd_k = 128$.
The pairwise interaction matrix $U$ is learned by applying a 4-layer pointwise MLP with (64, 64, 64, 8) channels and GELU nonlinearity from the 4 features motivated by Ref.~\cite{Dreyer:2020brq}, $(\ln \Delta, \ln k_{\text{T}}, \ln z, \ln m^2)$, where
\begin{align}
\begin{split}
    \Delta &= \sqrt{(y_a - y_b)^2 + (\phi_a - \phi_b)^2}, \\
    k_{\text{T}} &= \min(p_{\text{T},a}, p_{\text{T},b}) \Delta, \\
    z &= \min(p_{\text{T},a}, p_{\text{T},b}) / (p_{\text{T},a} + p_{\text{T},b}), \\
    m^2 &= (E_a+E_b)^2 - \|\mathbf{p}_{a}+\mathbf{p}_{b}\|^2,
\label{eq:interaction}
\end{split}
\end{align}
$y_i$ is the rapidity, $\phi_i$ is the azimuthal angle, $p_{\text{T},i} = (p_{x, i}^2+p_{y, i}^2)^{1/2}$ is the transverse momentum, and $\mathbf{p}_i=(p_{x,i}, p_{y,i}, p_{z,i})$ is the momentum 3-vector and $\|\cdot\|$ is the norm, for $i=a$, $b$. We use the exact pre-trained model weights provided from the Particle Transformer repository ~\cite{Qu:2022mxj}.

\vspace{-0.5em}
\section{Results}
\vspace{-0.5em}

\begin{figure}[ht]
    \centering
        \includegraphics[width=0.32\textwidth]{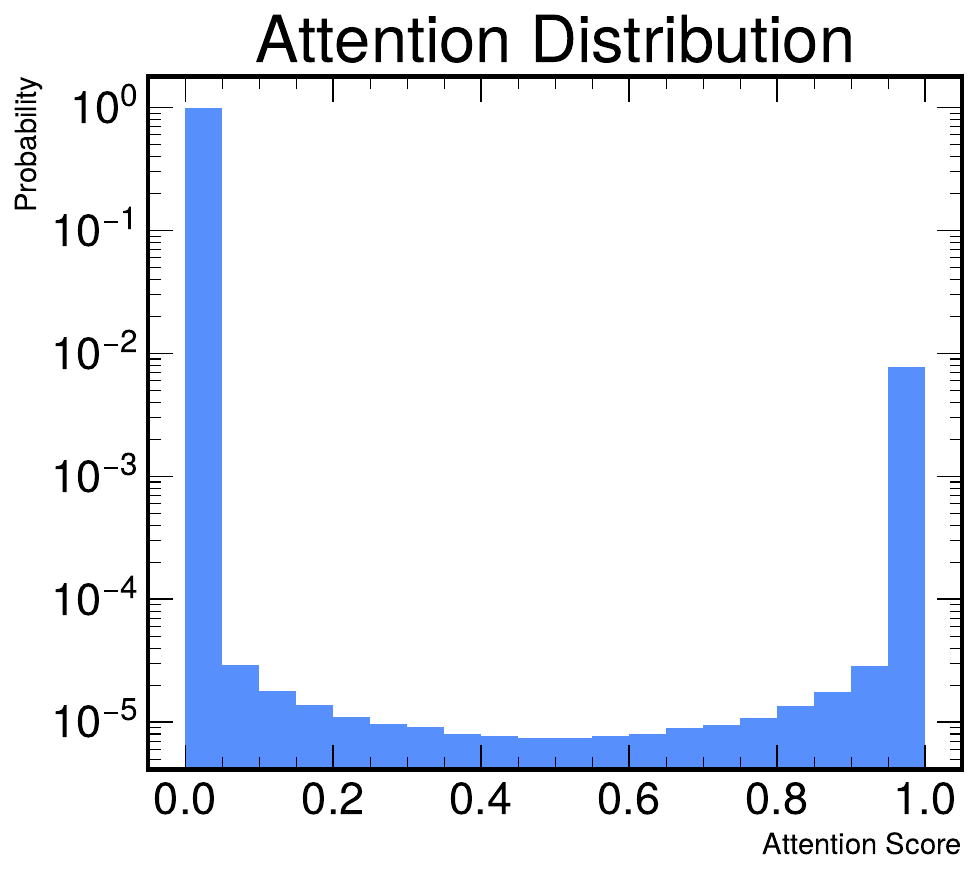}  
        \includegraphics[width=0.32\textwidth]{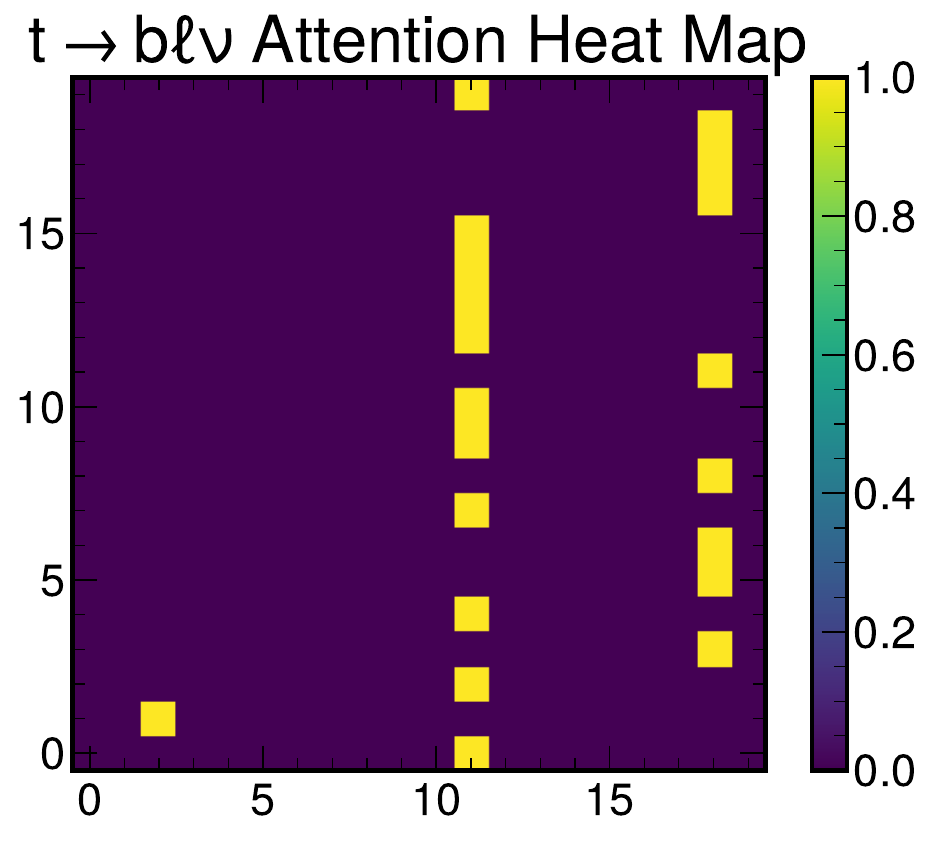}  
        \includegraphics[width=0.32\textwidth]{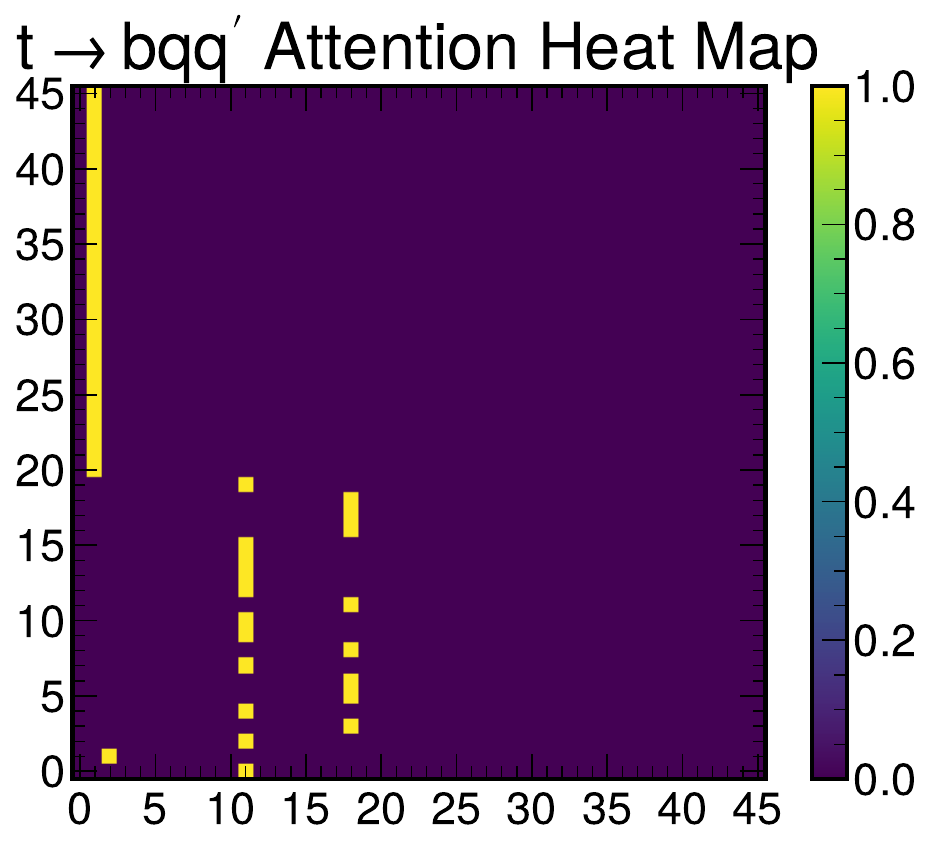}  
    \caption{Distribution of attention scores (left) illustrating the binary nature.
    Heat map of attention scores for example jets in the \tblv (center) and \tbqq classes (right) between the constituent particles.}
    \label{fig:heatmap}
\end{figure}

\paragraph{Distribution of Attention Scores} We visualize the attention scores across all heads in a heat map and observe that the ParT model's attention scores exhibit a nearly binary (on/off) pattern.
As shown in Fig.~\ref{fig:heatmap}, the distribution of attention scores reveals that most values are either close to one or zero.
This binary characteristic indicates that particles generally attend to, at most one other particle. 
This behavior raises the intriguing question of whether the model is capable of capturing underlying physical laws and how we might interpret the specific physics the model is learning from the data.
The binary nature of the attention weights also suggests potential for more efficient transformer models by reducing the computational complexity of attention mechanisms, focusing only on key interactions between particles.

\begin{figure}[ht]
    \centering

    \begin{tabular}{ccc}
        \textbf{$t \to b\ell\nu$} & \textbf{$t \to bqq'$} & \textbf{$H \to 4q$} \\
        \includegraphics[width=0.32\textwidth]{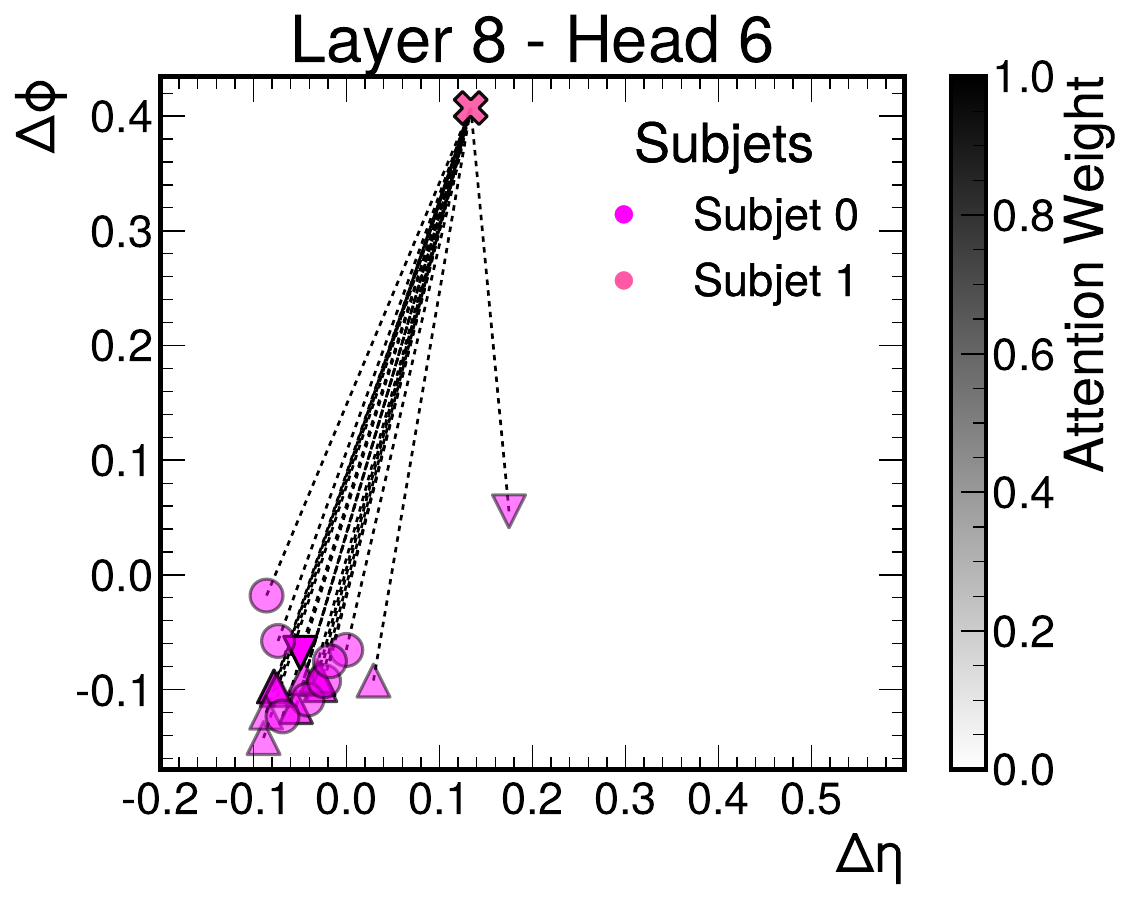} &
        \includegraphics[width=0.32\textwidth]{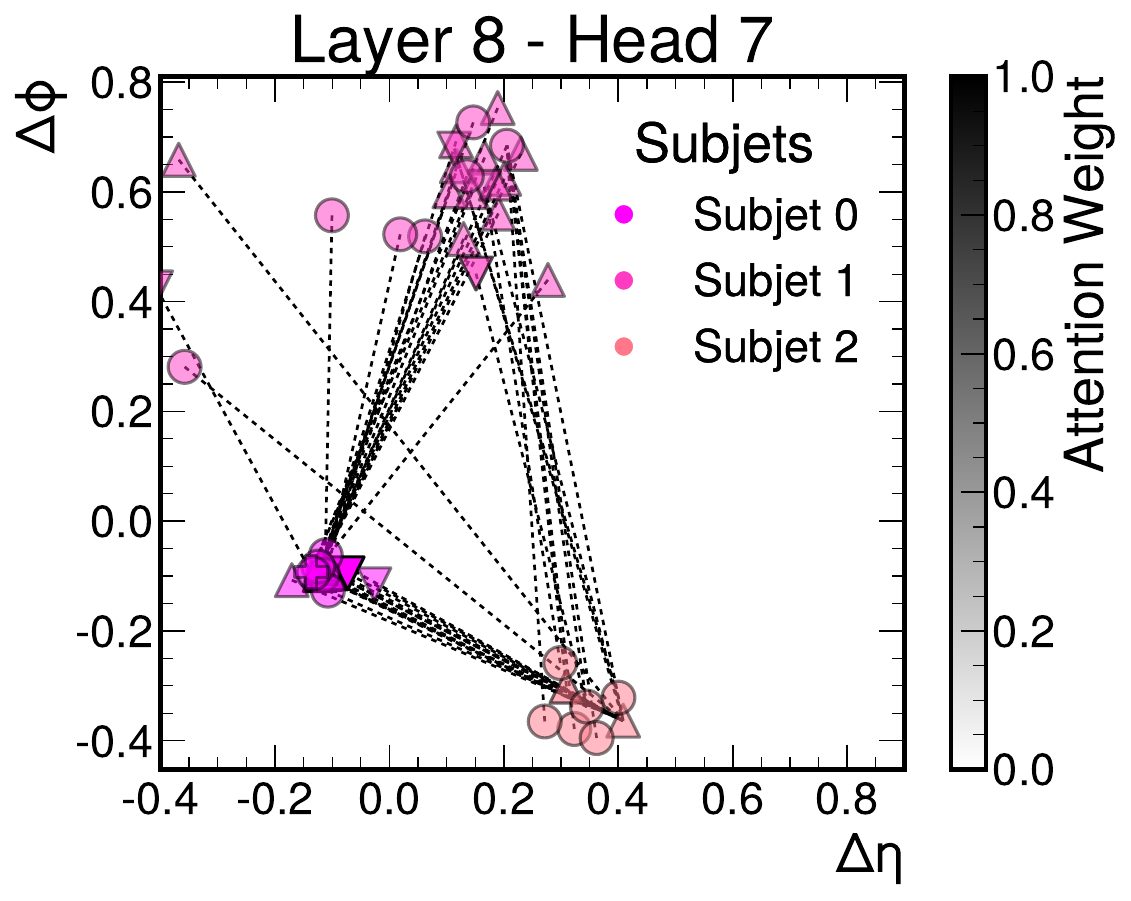} &
        \includegraphics[width=0.32\textwidth]{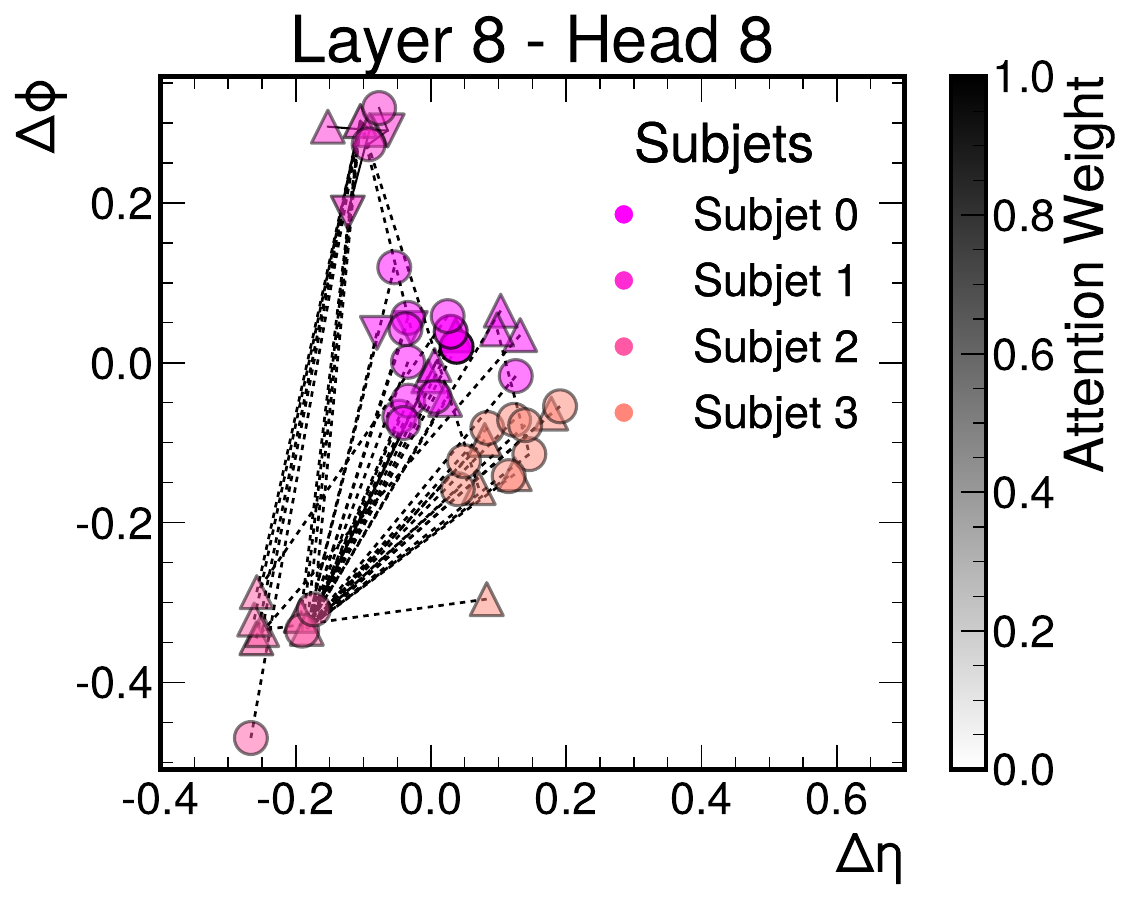} \\
        \includegraphics[width=0.32\textwidth]{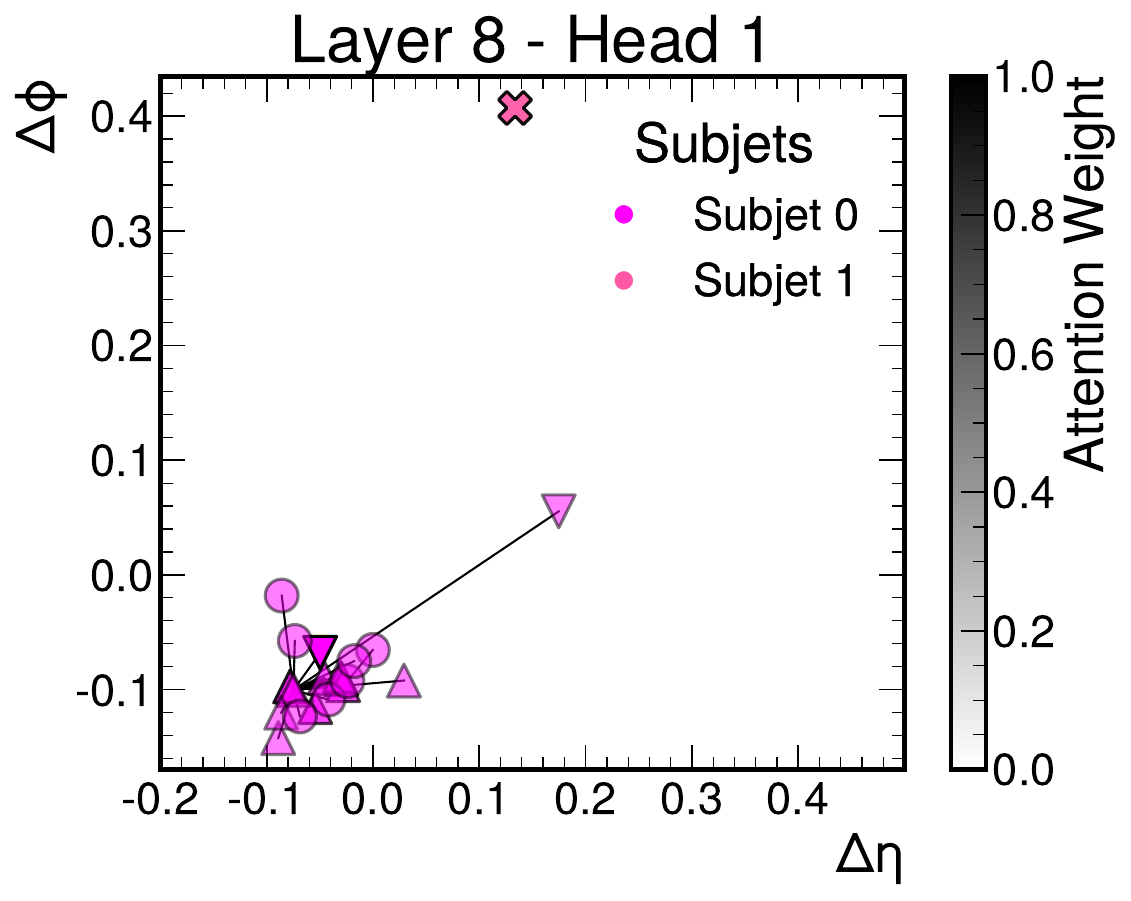} &
        \includegraphics[width=0.32\textwidth]{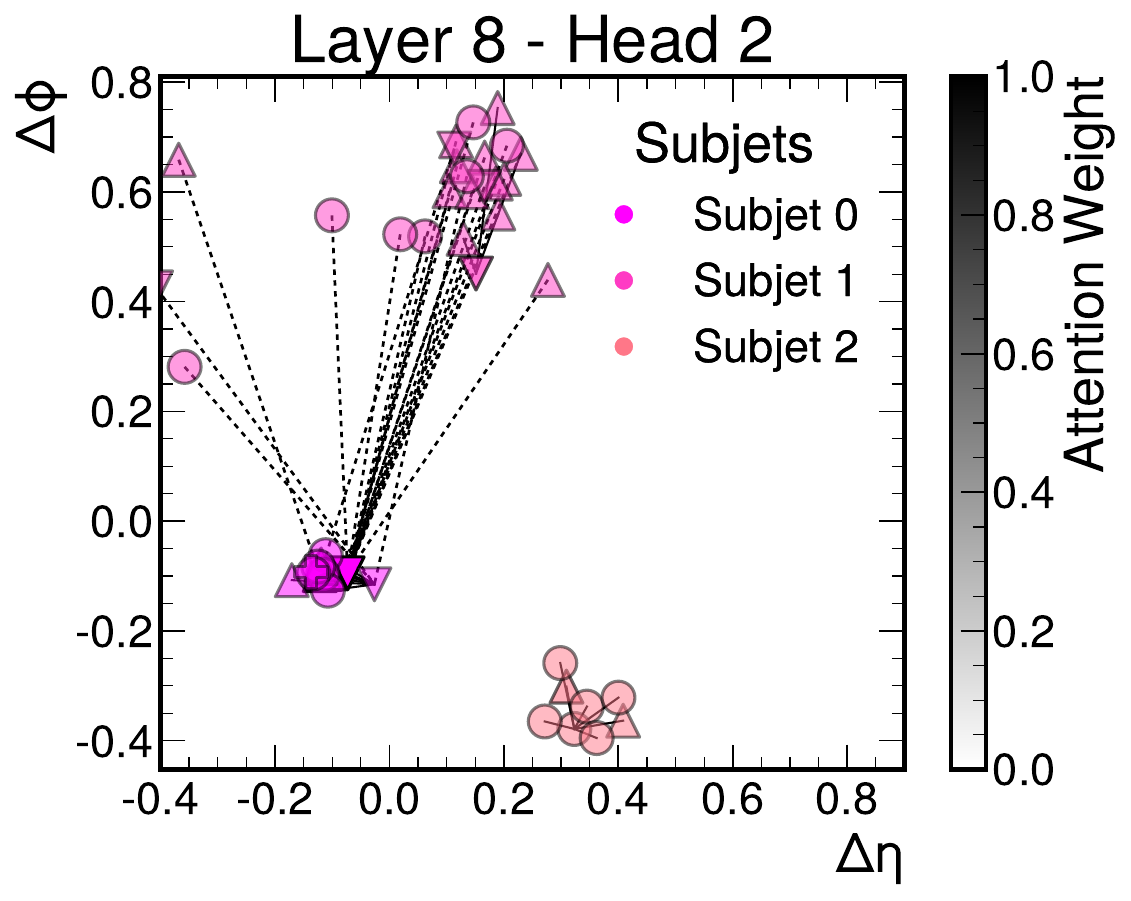} &
        \includegraphics[width=0.32\textwidth]{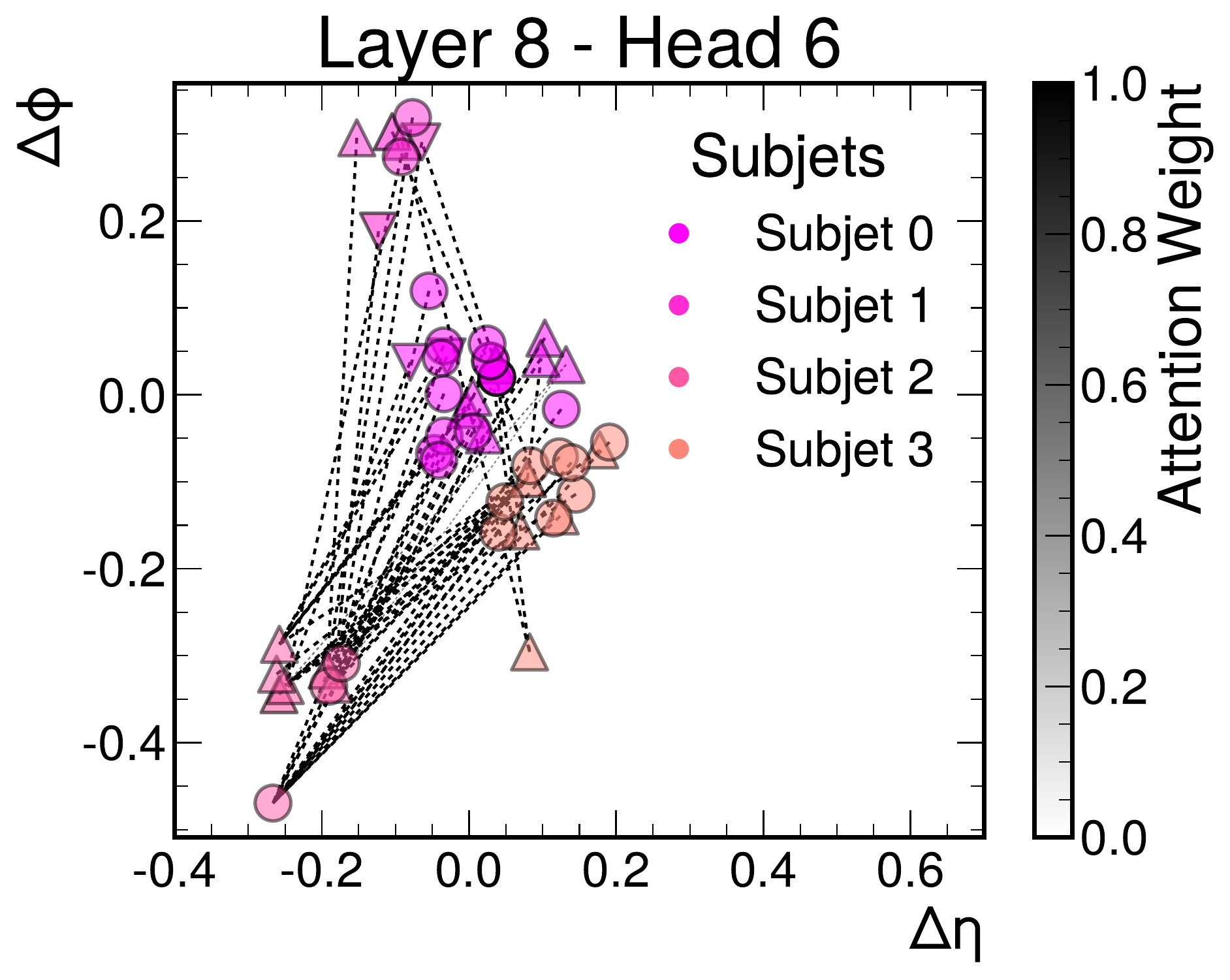} \\
        \includegraphics[width=0.32\textwidth]{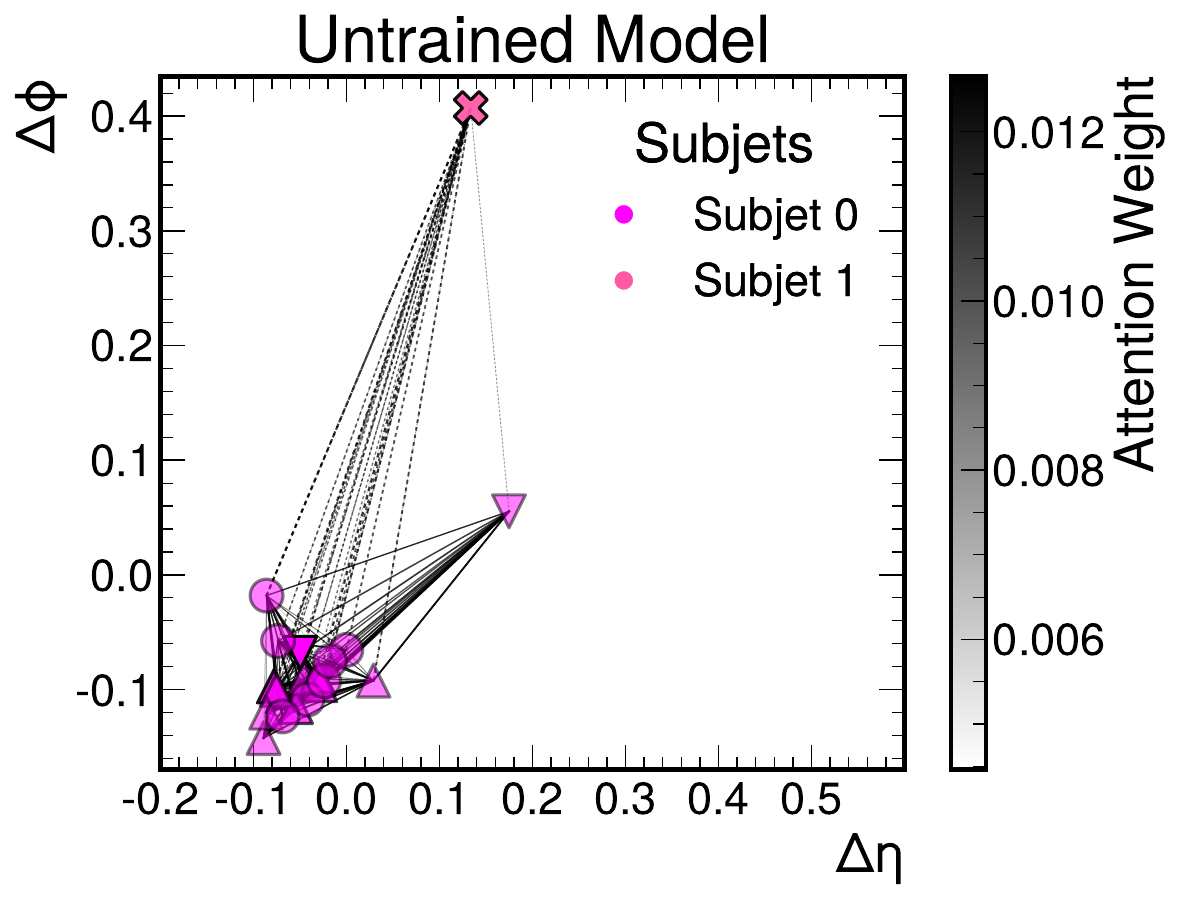} &
        \includegraphics[width=0.32\textwidth]{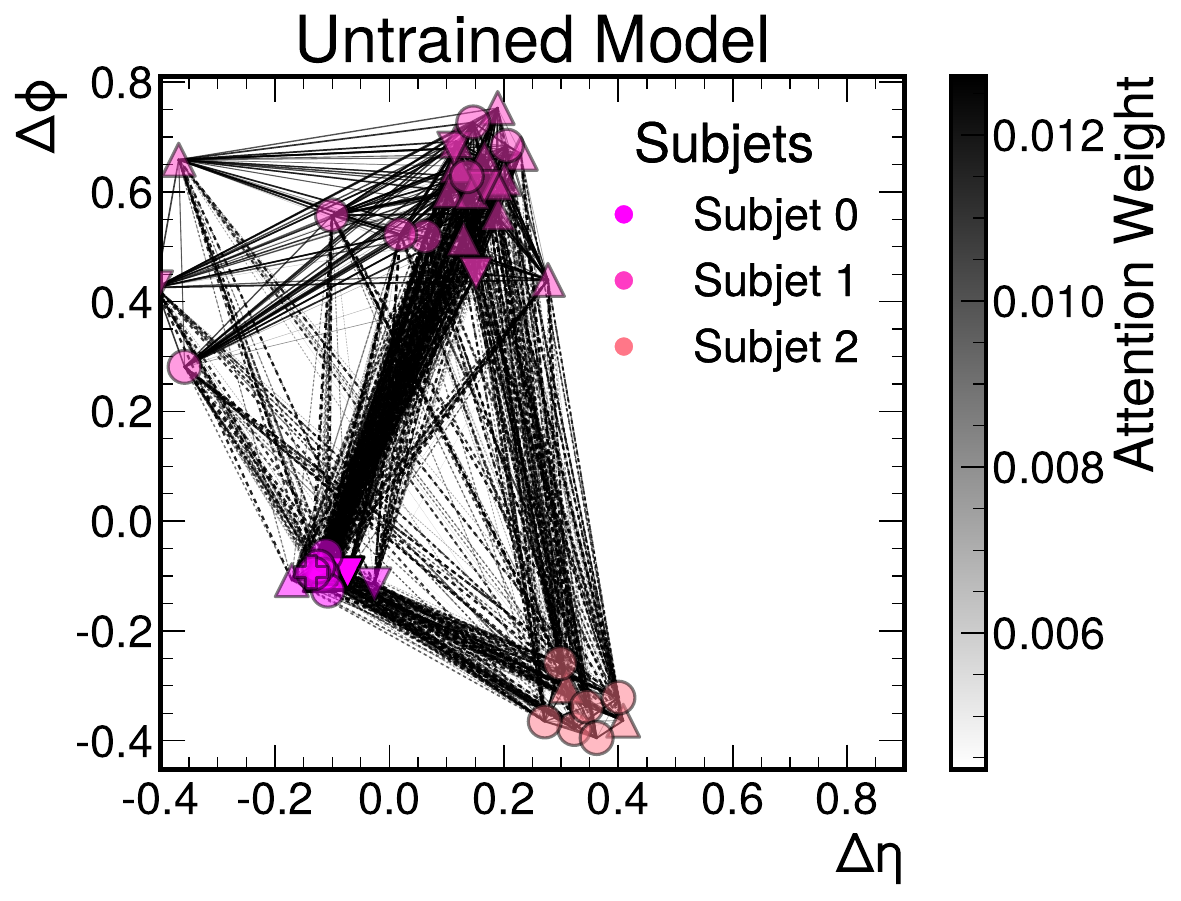} &
        \includegraphics[width=0.32\textwidth]{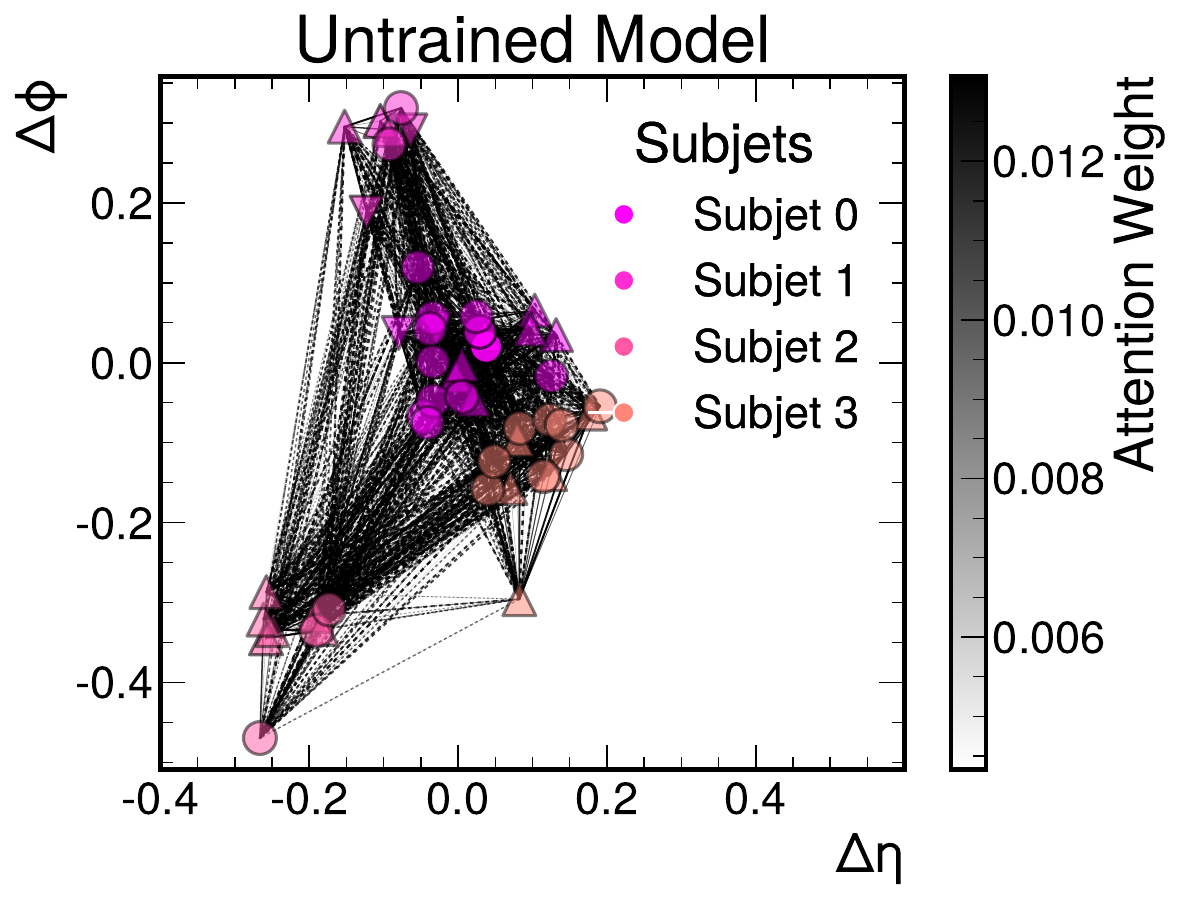} \\
    \end{tabular}

    \caption{Visualization of attention values in the $\eta$-$\phi$ plane for pairs of particles (\ding{54}: muon, \ding{115}: charged hadron, \ding{116}: neutral hadron, \ding{108}: photon, and \ding{58}: electron) within jets. The transparency of each point is proportional to the transverse momentum of the particle ($\pt$). The intensity of the connecting lines represents the magnitude of the attention scores. Solid lines depict intra-subjet connections, while dotted lines represent inter-subjet connections. The plots in the first column correspond to $t \to b\ell\nu$, the second column to $t \to bqq'$, and the third column to $H \to 4q$ processes.}

    \label{fig:ep_maps}
\end{figure}

\paragraph{Particle Attention Graphs} We decluster each jet into a specific number of subjets using the $k_\text{T}$ algorithm~\cite{Catani:1993hr} implemented in the FastJet package~\cite{Cacciari:2011ma} with Python bindings~\cite{Roy:2022rlt}.
We cluster into two subjets for leptonic top quark decays (\tblv), three subjets for hadronic top decays (\tbqq), and four subjets for Higgs decays into four quarks (\hqqqq).
For each process, we visualize the attention scores by representing each particle as a point in $\eta$-$\phi$ space in Fig.~\ref{fig:ep_maps}.
We compare these results with those obtained with randomly initialized attention weights.
We focus on the heads in the final layer of P-MHA. 

In Fig.~\ref{fig:ep_maps}, we observe that compared to an untrained, randomly initialized model, ParT demonstrates a significantly stronger tendency to attend to the lepton in the \tblv process.
This behavior indicates that ParT recognizes the importance of the lepton in classifying the underlying resonance and decay mode.
Additionally, in \tbqq and \hqqqq P-MHA captures both inter-subjet and intra-subjet interactions.
Specifically, ParT exhibits attention between different subjets and also within individual subjets. 
For example, in Fig.~\ref{fig:ep_maps}, P-MHA connects all different subjets but does not form connections within the subjets.
In contrast, P-MHA focuses its attention on interactions between subjets 0 and 1 while neglecting connections with subjet 2, instead of forming connections within subjet 2.

\begin{figure}[h]
    \centering

    \includegraphics[width=0.31\textwidth]{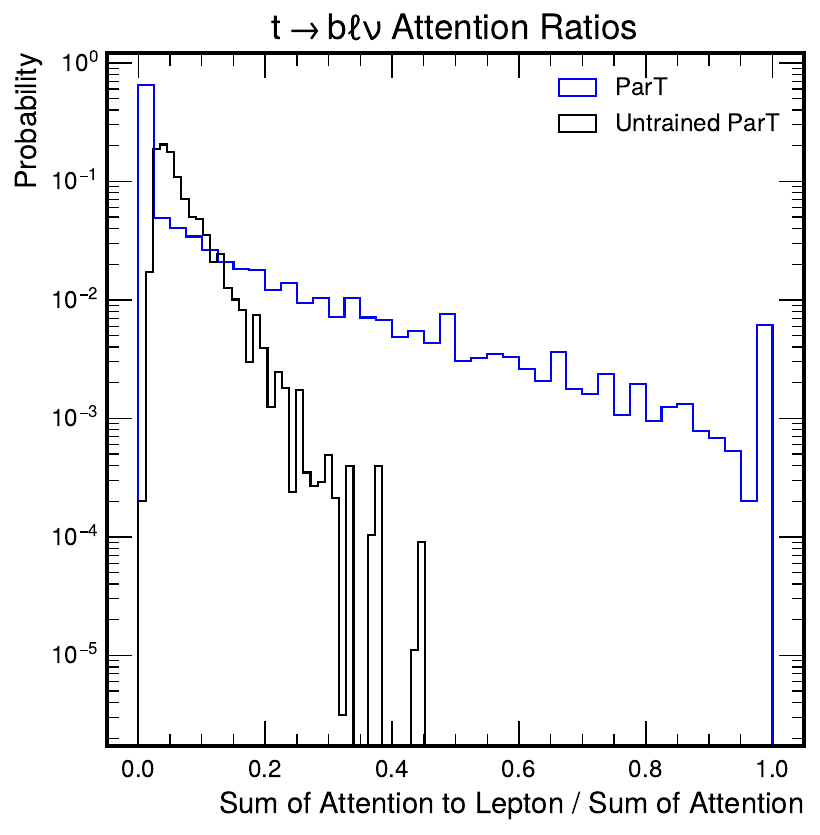} \includegraphics[width=0.31\textwidth]{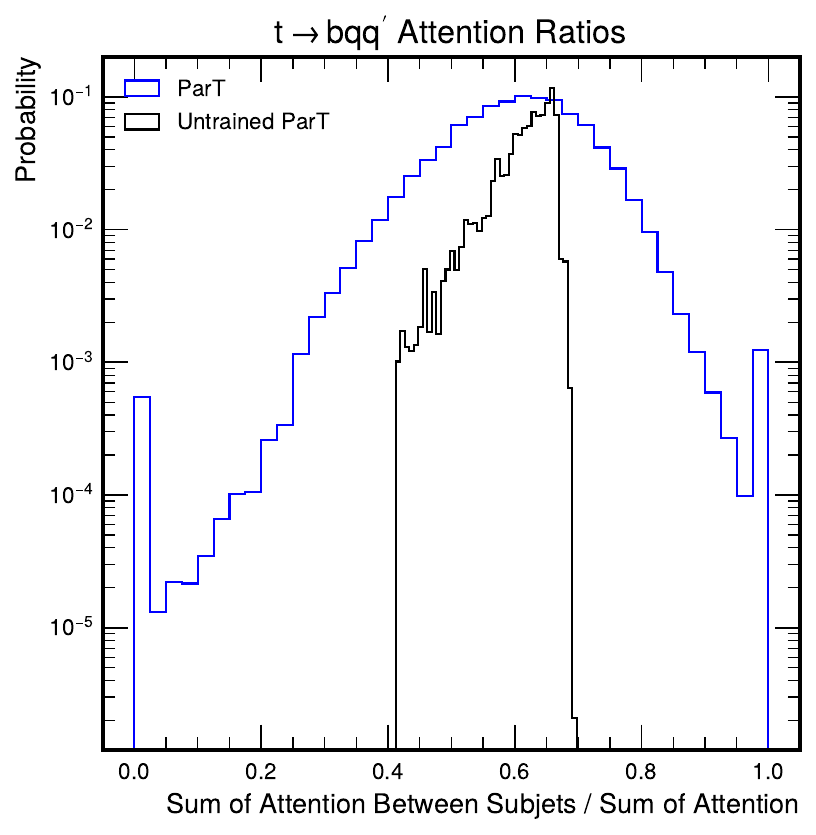} \includegraphics[width=0.31\textwidth]{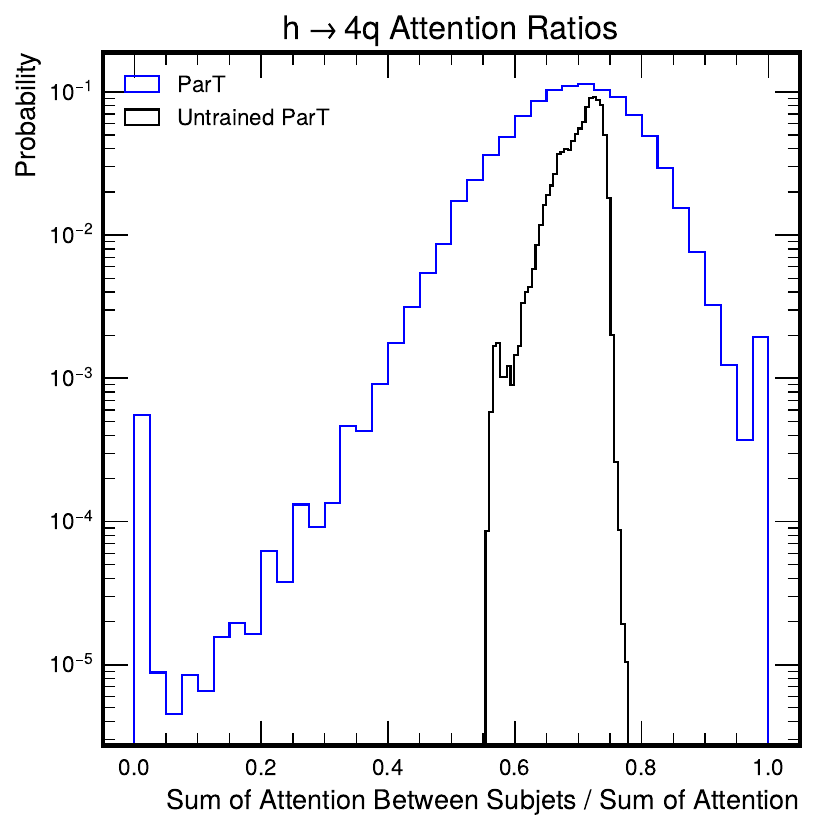}
   \caption{Distribution of the proportion of attention values that attend to leptons in \tblv (left). Distribution of the proportion of attention in between subjets in the \tbqq (center) and \hqqqq classes (right)}
   \label{fig:sumAtt}

\end{figure}

In order to quantify how often ParT exhibits these behaviors, we display the distribution of attention values for different jet classes.  
In particular, we find the sum of all attention values to a lepton in \tblv jets, and divide it by the sum of all attention values in the head.
Since the sum of all attention values in a head should be equal to the number of particles, this effectively is the proportion of attention weights that attend to the lepton.
Similarly, for \tbqq and \hqqqq jets, we display the sum of attention weights between particles in different subjets divided by the sum of all attention weights.
In Fig.~\ref{fig:sumAtt}, we observe that the trained ParT model is more likely to have a head that consists of only inter-subjet connections or only intra-subjet connections than an untrained, randomly initialized model.
ParT attending to only leptons or to no leptons for \tblv jets, and only intra-subjet or only inter-subjet for \tbqq and \hqqqq jets is some evidence that ParT is learning relevant physical properties of the jet data. 

\paragraph{Limitations}
Our analysis is constrained to the final attention layer, limiting insights into how earlier layers contribute to the representations. Additionally, the observed attention patterns may depend on the specific subjet clustering algorithm used, which could influence the model's interpretability. Exploring intermediate layers and varying subjet algorithms in future studies could provide a more comprehensive understanding of ParT's attention mechanisms.
\vspace{-0.5em}
\section{Optimizing Attention 
Mechanism for Efficiency}
\vspace{-0.5em}
\begin{table}[ht]
\caption{Performance of the ParT model with each attention head constrained to use a maximum of $k$ particles.}
\centering
\resizebox{\linewidth}{!}{%
\begin{tabular}{cccccccccccc}
\hline
\toprule
                & \multicolumn{2}{c}{}   & \hbb      & \hcc      & \hgg      & \hqqqq    & \hlvqq    & \tbqq     & \tblv     & \wqq      & \zqq      \\
                Max particles & Accuracy  & AUC       &\rej{50\%} &\rej{50\%} &\rej{50\%} &\rej{50\%} &\rej{99\%} &\rej{50\%}&\rej{99.5\%}&\rej{50\%} &\rej{50\%} \\
\hline
1     & 0.770 & 0.9754 & 4396 & 1325 & 58 & 246 & 1944 & 2384 & 6601 & 217 & 175 \\
2    & 0.798 & 0.9795 & 5780 & 1862 & 67 & 390 & 2894 & 4219 & 10050 & 286 & 231 \\
3    & 0.814 & 0.9817 & 6757 & 2268 & 73 & 515 & 3460 & 5917 & 11111 & 341 & 272 \\
4    & 0.825 & 0.9831 & 7326 & 2558 & 79 & 661 & 3884 & 7813 & 12422 & 379 & 299 \\
6    & 0.838 & 0.9849 & 8621 & 3063 & 89 & 938 & 4474 & 12821 & 13889 & 438 & 339 \\
10   & 0.851 & 0.9865 & 9852 & 3670 & 103 & 1352 & 5102 & 20619 & 15748 & 502 & 375 \\
20   & 0.859 & 0.9875 & 11111 & 4237 & 118 & 1802 & 5263 & 29412 & 16000 & 539 & 399 \\
30   & 0.860 & 0.9877 & 10989 & 4193 & 122 & 1859 & 5362 & 33898 & 16129 & 543 & 402 \\
128           & 0.861 & 0.9877 & 10638 & 4149 & 123 & 1864 & 5479 & 32787 & 15873 & 543 & 402 \\
\hline
\end{tabular}
}
\vspace{0.5em} 

\label{tab:numParticles}
\end{table}

Leveraging the binary nature of P-MHA, we constrain the total number of particles used in the attention mechanism, and explore the impact of the number of particles attended to in each attention head on the overall performance.
For only one particle, we directly replace attention with softmax.
For other numbers of particles, we keep the top-$k$ highest attention particles, setting the rest of the particles to zero, and then applying softmax over those $k$ particles (out of 128 maximum).
We find that with just 1 particle, some performance is maintained and with 30 particles, the performance of ParT is recovered as shown in Table \ref{tab:numParticles}.
By filtering the particles that P-MHA can consider, we can potentially reduce the number of calculations and increase computational efficiency.

\vspace{-0.5em}
\section{Summary and Outlook}
\vspace{-0.5em}
The ParT model demonstrates a unique binary attention pattern, in which each particle typically attends to at most one other particle.
This focused attention mechanism contrasts with other transformer models, such as vision transformers, which spread attention across many elements. 
The sparse and selective attention in ParT prioritizes key relationships, potentially uncovering important substructures in jet tagging, and enhancing model interpretability in high-energy physics.

This binary pattern also opens up opportunities for optimization.
This could lead to more efficient architectures without compromising performance, particularly in tasks such as jet tagging, where the identification of essential interactions is vital.
Future work could explore these optimizations to further improve both the interpretability and efficiency of transformers in physics applications. This work will be helpful in designing a more efficient physics-based attention mechanisims.

In addition, the importance of the interaction matrix for the performance of ParT has not been explored fully.
In Ref.~\cite{wu2024jettaggingmoreinteractionparticle}, many P-MHA layers are replaced with the interaction matrix and find improved performance and efficiency.
Understanding what the interaction matrix is learning to augment P-MHA could lead to a more efficient architecture. 

\paragraph{Broader Impact}
Studies that interpret machine learning algorithms are important to increase confidence in model predictions and also to improve the performance of future machine learning models.
This study uses an xAI method in order to explain the performance of the state-of-the-art transformer model for the LHC, by demonstrating that the model is indeed learning physics.
This study opens doors for future studies to extract and discover physics information in machine learning models that are currently not being used for prediction. 

The code used for this analysis is publicly available at: \url{https://github.com/aaronw5/Interpreting-Transformers-for-Jet-Tagging}

\acksection
\vspace{-0.5em}
This work is supported by the DOE Office of Science, Award No. DE-SC0023524, Fermi Research Alliance, LLC under Contract No. DE-AC02-07CH11359 with the DOE, Office of Science, Office of High Energy Physics, LDRD L2024-066-1, DOE Office of Science, Office of High Energy Physics ``Designing efficient edge AI with physics phenomena'' Project (DE-FOA-0002705), DOE Office of Science, Office of Advanced Scientific Computing Research under the ``Real-time Data Reduction Codesign at the Extreme Edge for Science'' Project (DE-FOA-0002501),a nd DOE Early Career Research Program Award No. DE-SC0021187.
The work is also supported in part by RCSA Grant \#CS-CSA-2023-109, Sloan Foundation Grant \#FG-2023-20452, NSF awards CNS-1730158, ACI-1540112, ACI-1541349, OAC-1826967, OAC-2112167, CNS-2100237, CNS-2120019, the University of California Office of the President, and the University of California San Diego's California Institute for Telecommunications and Information Technology/Qualcomm Institute, AI2050 program at Schmidt Futures (Grant G-23-64934), and the NSF HDR Institute A3D3 (PHY-2117997).


\bibliography{reference}

\end{document}